# Optical Properties and Behavior of Whispering Gallery Mode Resonators in Complex Microsphere Configurations: Insights for Sensing and Information Processing Applications

Yaser M. Banad[1,#], Syed Mohammad Abid Hasan[2,#], Sarah S. Sharif[1,3*], Georgios Veronis[4,5*], and Manas Ranjan Gartia[2,*]

*Abstract*- **Whispering gallery mode (WGM) resonators are garnering significant attention due to their unique characteristics and remarkable properties. When integrated with optical sensing and processing technology, WGM resonators offer numerous advantages, including compact size, high sensitivity, rapid response, and tunability. This paper comprehensively investigates the optical properties and behavior of WGMs in complex microsphere resonator configurations. The findings underscore the potential of WGMs in sensing applications and their role in advancing future optical information processing. The study explores the impact of configuration, size, excitation, polarization, and coupling effects on the WGMs' properties. The paper provides crucial insights and valuable guidance for designing and optimizing microsphere resonator systems, enabling their realization for practical applications.**

*Keywords: whispering gallery modes, microsphere resonators, optical information processing*.

## I. INTRODUCTION

Recent progress in the field of nanolasers includes the invention of microdisk 1, photonic crystal 2, and nanowire lasers 3. Recent developments in the field of nanostructured surfaces 4-6 led researchers to novel nanolaser structures 7, such as metal-based-coated nanolasers 8-10 and nanoparticle lasers 11, 12; plasmonic-based-nanowire 13, 14, photonic crystal 15, 16 and waveguide embedded nanolasers 17; nanopatch and nanodisk lasers; coaxial and pseudo-wedge nanolasers 18. Diffraction-limited lasers with sizes of approximately half the lasing wavelength are widely used in these nanoscale designs. These lasers can be used as coherent light sources in optoelectronic circuits 19 and for cellular imaging 20. They can sometimes be used to deliver useful information related to light-matter interactions for fluorescence, photocatalysis, quantum processes, light-dependent oscillations of electrons in metal-based nanostructures, and nonlinear optical processes 11, 21-24. The miniaturization of these lasers, including designs based on plasmonic crystals and semiconductor nanowires, may lead to new applications in sensing and switching 25. One of the promising aspects of semiconductor-based nanowires is to serve as both gain media and resonance cavities for light amplification 26-28. Further, including whispering gallery modes (WGMs) in such designs can introduce nonlinear effects within the resonator, which could lead to new applications for activation functions in optical neural networks 29.

In addition, the importance of optical bio-probes in biomedical engineering is immense, since fluorescent proteins, plasmonic nanoparticles, and dyes as optical readers help to understand many biological mechanisms. Optical bio-chemical sensors utilize interactions between electromagnetic waves and biological molecules to detect changes in output signals and acquire the necessary concentration information. However, the broad emission spectra of fluorescent and luminescent probes often create issues in applying these techniques for high-resolution imaging and ultra-sensitive bio-sensing 30. WGM-based resonators with microsphere structures help to overcome these issues by offering narrow linewidth. WGM resonator-based optical biochemical sensors have emerged as a strong contender for various bio-chemical sensing applications due to their high-quality Q factor, high sensitivity, small mode volume, rapid response, and ease of integration 31. WGM-based lasers are also useful for optical gain doping and proper pumping by using low threshold lasing 32 to overcome compliant dosing of the resonator along with the laser species 33, 34. These lasing operations can be tuned in various ways through geometry or the refractive index. The resonant modes can also be manipulated using polymer droplets with different sizes 35, 36. Ausman and Schatz found that the surface-enhanced Raman scattering enhancement factors for a molecule vary inversely with respect to the resonance width, which shows the influence of WGM mode number and order 37.

Recently, WGM-based photoemission and lasing were accomplished using organic and inorganic microdisks and spheres 36, 38-45. WGM-based sensors were successfully

Y. M. Banad is with the School of Electrical and Computer Engineering, University of Oklahoma, Norman, OK 73019, U.S.A.; S. M. A. Hasan is with the Department of Mechanical and Industrial Engineering, Louisiana State University, Baton Rouge, LA, 70803, U.S.A.; S. S. Sharif is with the School of Electrical and Computer Engineering, University of Oklahoma, Norman, OK 73019, U.S.A., and the Center for Quantum Research and Technology, University of Oklahoma, Norman, OK 73019, U.S.A.; G. Veronis is with the School of Electrical Engineering and Computer Science, Louisiana State University, Baton Rouge, LA, 70803, U.S.A., and Center for Computation and Technology, Louisiana State University, Baton Rouge, LA, 70803, U.S.A.; M. R. Gartia is with the Department of Mechanical and Industrial Engineering, Louisiana State University, Baton Rouge, LA, 70803, U.S.A. E-mails: *bana@ou.edu*; *abid.mech.07@hotmail.com*; *s.sh@ou.edu*; *gveronis@lsu.edu*; *mgartia@lsu.edu*; # Equal contribution authors; *Corresponding authors.

demonstrated for bio-sensing applications 45-47, to detect label-free DNA 48, lead ion 49, glucose concentration 50, and real-time detection of protein secretion from living cells 51. In addition, intensifying the optical signals using WGMs can be crucial in creating an optical neural network 29 as higher-intensity WGMs allow for stronger and more robust optical signals, enabling efficient transmission and processing of data for optical systems on a chip.

In this paper, we present an optoplasmonic amplifier operating within the visible range, which demonstrates the capability of generating an internal Raman signal through injection seeding. Our study explores the utilization of microspheres arranged in various configurations, including single spheres, two spheres with equal or unequal sizes, three spheres with equal sizes and multiple spheres with different sizes and configurations. We conduct an in-depth analysis of the impact of excitation and polarization with respect to different sphere configurations and excitation locations. Notably, we observe a mode position shift corresponding to variations in microsphere sizes. Through this comprehensive investigation, our work provides profound insights into the behavior of WGMs, which hold broad potential applications in bio-chemical sensors and optical information processing.

## II. EXPERIMENTAL SETUP AND THEORETICAL PRINCIPLES

Figure 1 displays scanning electron micrographs (SEMs) illustrating various microsphere configurations and the corresponding experimental setup. We conducted experiments to investigate the influence of WGMs in diverse arrangements. These configurations included linear arrangements, variations in angles, sizes, and quantities of microspheres. We carefully selected microspheres with diameter of 10.1 μm and arranged them in direct contact on a glass plate. The biotinylated microspheres are conjugated to NeutrAvidin Dylight 650 dye. All spectroscopic measurements were conducted under continuous wave (CW) conditions using a continuous wave HeNe laser with a 632.8 nm wavelength and a numerical aperture of 0.45, facilitated by a photoluminescence (PL) microscope. The microresonators were composed of Polystyrene with a refractive index of 1.58, positioned on a silicon nanopillar substrate. Through micromanipulation on a glass slide attached to a three-dimensional (3-D) micromanipulation stage, we assembled these pre-sorted spheres, each possessing nearly identical resonant characteristics, into specific molecular topologies. To secure the spheres during our experiments, we applied a thin epoxy layer (a few microns thick) to the surface of a glass plate.

In multi-sphere systems, such as two-sphere arrangements, we employed the laser as an excitation source for one sphere while the remaining spheres acted as passive receivers 52. Employing the concept of injection seeding with the narrow bandwidth radiation from Raman scattering of the pump, we devised an optically-pumped photonic/plasmonic amplifier, analogous to macroscopic master oscillator-power amplifiers (MOPAs) 53. This approach allowed us to effectively decouple the Q-factor from the WGM resonator. In other words, the amplifier's Q-factor was no longer limited by the WGM resonator but was instead determined by the pump laser linewidth and the Raman mode providing the seed radiation. The amplification resulted from the stimulated emission of photons during molecular transitions from a higher energy state to a lower energy state, initially populated by the pump source. Our gain medium consisted of fluorescence molecules tethered to the exterior of the microsphere resonators, separated by a protein layer and a plasmonic surface. This tethering ensured that the entire gain medium remained within the evanescent field associated with the power circulating in the microresonator, thereby affecting both the effective lifetime of the excited singlet state and the stored energy within the medium. The frame was sealed at the bottom using a microscope slide and then filled with distilled water. To maintain a sufficiently high Q-factor, larger diameter spheres were preferred due to the reduced index contrast in water 54, 55.

Finite-difference time-domain (FDTD) simulations were performed (Lumerical FDTD Solutions). The 3-D FDTD method was used to calculate the field enhancement and field profiles in the microsphere resonators. The mesh size was chosen as 0.25 nm throughout the simulation domain, which was terminated to perfectly matched layers (PMLs) in all directions 51. WGMs occur when light is trapped in a microsphere by total internal reflection. An initial simulation using a dipole source was performed to locate the resonant frequencies of the structures. The wavelength range of the dipole sources was set to be 640-750 nm in order to excite the resonant frequencies in this range. The wavelength range covers the azimuthal mode numbers 60 – 71 for both TE and TM modes (See Supporting Information Figures S1). The sources were placed near the edge of the structures because modal fields in this region are strong. The Fourier transform of the field over time was calculated at a monitor point to find the resonant frequencies of the structure. Frequency domain profile monitors were set at these frequencies to record the field profiles.

In a microresonator configuration, light rays propagate in the form of WGMs, which results from trapping the rays within the microresonator by almost total internal reflection from the curved surface of the resonator. The approximate ray propagation path within the microresonator can be conceptualized as a polygon. When longer resonant wavelengths are present, the polygon has fewer edges. In general, if the radius of the WGM resonator R is much larger than the resonant wavelength λ, i.e. R >> λ, and the ray strikes the surface at near-glancing incidence, the resonance condition occurs when the light path length of the trapped ray is approximately equal to a multiple of the resonant wavelength: $2\pi R n_{eff} = m\lambda$, where, $n_{eff}$ represents the effective refractive index of the optical mode, and m is an integer representing the WGM angular momentum. Hence, any wavelengths that satisfy this equation will be resonant wavelengths of the WGM resonator. Moreover, when the effective refractive index $n_{eff}$ changes, the resonant wavelengths will shift accordingly. While the optical WGM remains confined within the resonator, some of its energy will extend into the surrounding medium as an evanescent field. Environmental variations within this evanescent field region will affect the modes of the WGM, resulting in shifts of the resonance modes, widening mode linewidth, and the spectral separation of previously degenerate modes. This principle forms the basis of sensing in WGM resonator-based sensors—the resonant wavelength of the WGM resonator shifts due to variations in the effective refractive index.

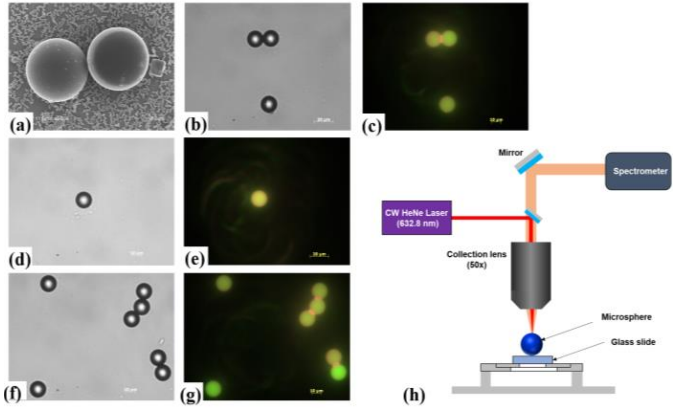

Figure 1. (a) SEM image of a two-microsphere structure with 10.1 μm microsphere diameter. (b) Optical, and (c) photoluminescence (PL) image of the two-microsphere structure. (d) Optical, and (e) PL image of a single microsphere system. (f) Optical, and (g) PL images of a three-microsphere resonator system. (h) Schematic showing the experimental setup.

### III. RESULTS AND DISCUSSION

This section presents the results and discussion of microspheres arranged in various configurations, including single spheres, two spheres with equal or unequal sizes, three spheres with equal sizes, and multiple spheres with different sizes and geometrical arrangements. We comprehensively analyze the influence of excitation, polarization, sphere properties, and excitation positions.

Figure 2 shows the results for a single microsphere resonator. Figure 2(a) displays a schematic, bright field, and fluorescence image of a 10 μm diameter microsphere resonator, which is significantly larger than the resonant wavelength of the WGMs. Figure 2(b) depicts the electric field intensity distribution for a microsphere with the same diameter. WGMs are surface modes supported by the dielectric microresonators. We observe trapping of light within the spherical resonator, resulting in a highly intensified electric field at a wavelength of 675 nm. The light propagates in the form of WGMs, which exist due to the trapping of light rays through total internal reflection from the curved surface of the resonator. Our simulations demonstrate that WGMs exhibit a pronounced concentration of light along the edges of the sphere, leading to significantly enhanced electric field intensity at this specific resonance wavelength. This strong local electric field enhancement and high confinement could contribute to improved sensitivity and detection capabilities of WGM sensors. Furthermore, the fluorescence spectra emanating from the dye anchored to the microsphere significantly bolsters the optical gain of the resonator amplifier, particularly at the wavelength of 675 nm. When a signal is injected into the amplifier, it extracts energy stored within the gain profile, assuming the gain medium is homogeneously broadened, and the injection seed wavelength aligns with the dye gain spectrum. Notably, the PL spectra exhibits varying intensities at different laser power levels, a characteristic attributed to the inherent nonlinearity within this optical system.

Figures 2(c)-2(e) display the relative intensity versus wavelength, represented by the PL spectra, within the 640-760 nm wavelength range for an amplifier incorporating a single-sphere resonator. The corresponding TE TM modes are shown in Figure S2 and Table S1. The experiments were conducted with three different laser power levels: 0.2 mW, 1 mW, and 2 mW. The corresponding Q-factors of WGM peaks, which are a measure of energy confinement, were measured. At higher laser powers of 2 mW and 1 mW, distinct strong peaks appear at 675 nm with Q-factors of 1370 and 1120, respectively. These dominant peaks exhibit approximately four times higher intensity compared to the subsequent highest Q-factors observed at the wavelength of 658 nm, resulting in high adjacent peak ratios for these two laser power levels. Conversely, at the lower laser power level of 0.2 mW, multiple short peaks emerge, excited across a wide range of wavelengths corresponding to various whispering gallery modes. Specifically, at this laser power level, 25 short peaks are observed with Q-factors exceeding 200 but not higher than 800. Additionally, the average peak-to-peak distance in wavelength of the optical signal at this laser power level is ~6 nm. The presence of a large number of WGM peaks with lower intensity can enhance the network's resilience to noise and signal degradation, creating optical signals that are less vulnerable to various forms of noise and attenuation. By utilizing highly intense and amplified WGMs at specific wavelengths, the network could mitigate the adverse effects of noise, enabling accurate and reliable data processing when used in future optical neural networks.

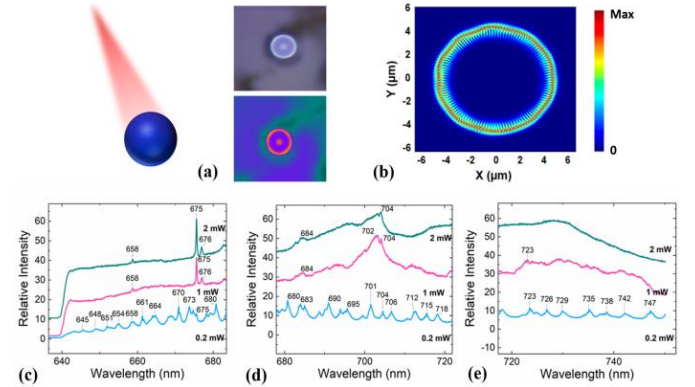

Figure 2. (a) A schematic, bright field, and fluorescence image for a single microsphere resonator. (b) Electric field intensity calculated for one sphere at λ = 675 nm and plotted in the xy plane (top view of the sphere). (c-e) PL spectra in the 640–750 nm wavelength range for an amplifier incorporating a 10.1 μm diameter single sphere resonator at excitation laser power levels of 0.2, 1, and 2 mW.

In Figure 3, we present the results obtained for a double microsphere resonator. Figure 3(a) shows a schematic and bright field image of a double microsphere resonator with equal sphere diameters of 10.1 μm. To gain insight into the coupling between photonic microsphere resonators, we conducted simulations to analyze the intensity distribution of the electric field in a double microsphere resonator with the same diameter at a specific wavelength of 667 nm, as depicted in Figure 3(b). Our simulations reveal that the WGMs effectively trap light within the first spherical resonator and light is also coupled to the adjacent spherical resonator. Similar to the single microsphere case, the WGMs exhibit a pronounced concentration of light along the edges of the spheres, leading to an enhanced electric field intensity at this particular resonance frequency. We observe that the electric field is highly intensified at the wavelength of 676 nm at the edges of the two spheres, while the light power trapped in the first microsphere resonator couples into the second microsphere resonator with a reduction in the electric field intensity observed at the points where the microsphere resonators make contact. This reduction in the electric field intensity is due to destructive interference at the contact point of two spheres.

Furthermore, Figure 3(c)-3(e) present the relative intensity versus wavelength of the PL spectra within the 650–750 nm wavelength range for an amplifier incorporating a double sphere resonator. As before, the experiments were conducted using laser power levels ranging from 0.2 mW to 1 mW until saturation. At a wavelength of 676 nm, the Q-factors were measured to be 1690 and 450 for laser power levels of 0.2 mW and 1 mW, respectively. For a laser power level of 1 mW, only one peak was observed at this wavelength, while at the lower laser power level of 0.2 mW, 15 short peaks of relative intensity emerged across a wide range of wavelengths corresponding to various WGMs. The Q-factors of these peaks did not exceed 400. The presence of a large number of WGM peaks with lower intensity is not desirable for many applications due to their vulnerability to various forms of noise and attenuation. By utilizing high field intensity WGMs at the 676 nm wavelength, the information encoded in the optical signal could effectively mitigate the adverse effects of noise, which could ensure accurate and reliable data processing in future optical neural networks.

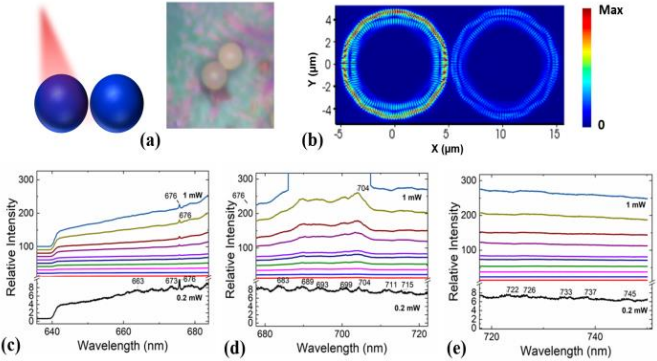

Figure 3. (a) A schematic and bright field image for a double microsphere resonator. (b) Electric field intensity calculated for two spheres at λ = 676 nm and plotted in the xy plane (top view of the spheres). (c)-(e) PL spectra in the 640–750 nm wavelength range for an amplifier incorporating a double sphere resonator with diameters of 10.1 µm.

Figure 4(a) shows a schematic and bright field image of a double microsphere resonator with unequal sphere sizes, where the smaller and larger microspheres have diameters of 10 µm and 13 µm, respectively. In Fig. 4(b), we depict the electric field intensity distribution for a similar unequal double microsphere resonator at the wavelength of 675 nm. The simulation results reveal that the WGMs exhibit a pronounced concentration of light along the edges of the smaller microsphere resonator, leading to a significant enhancement in electric field intensity at this resonance frequency. However, the light intensity is significantly smaller when it couples into the larger microsphere resonator. Figure 4(c) displays the PL spectra with various WGM peaks across the wavelength range of 640-750 nm for two cases: exciting the large or the small sphere. The PL spectra show that, when exciting the smaller sphere, there are 23 short resonance peaks with Q-factors between 50 to 100. However, when exciting the larger sphere, only two resonance peaks with Q-factors exceeding 100 and 9 other WGM peaks with Q-factors between 50 and 100 are observed. Figures 4(d)-4(f) show the spectral locations of these WGM peaks labeled with their corresponding wavelengths for these two cases of exciting the small and large spheres. At the wavelength of 675 nm, the dominant resonance peaks have the highest Q-factors of 600 and 450 when exciting the small and large spheres, respectively. In addition, the average peak-to-peak distance in wavelength of the resonance peaks are ~2.8 nm and ~5 nm for the cases of exciting the small and large spheres, respectively. These results demonstrate that the size of the microsphere resonator excited by the laser affects the optical properties of WGMs, leading to variations in the intensity and number of resonance peaks observed.

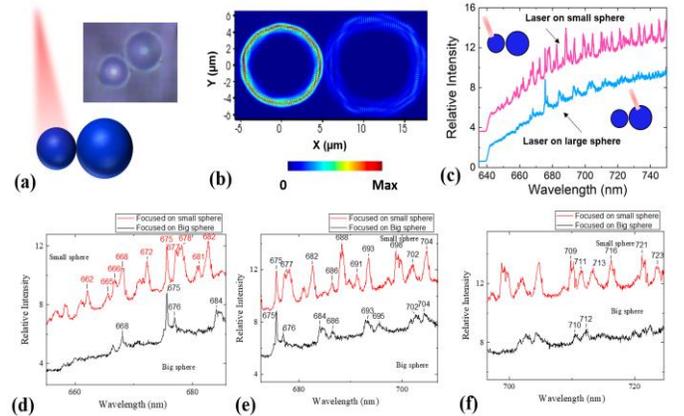

Figure 4. (a) A schematic and bright field image for a two microspheres resonator with unequal sphere sizes; (b) Electric field intensity calculated for the double microsphere resonator with unequal sphere sizes at 675 nm wavelength and plotted in the xy plane (top view of the spheres); (c) PL spectra in the 640–750 nm wavelength for an amplifier incorporating a two microsphere resonator with unequal sizes. (d-f) PL spectra with labeled WGM peaks ranging from 640 nm to 750 nm are shown for the amplifier with two microsphere resonators with unequal sphere sizes. Two excitation cases are depicted: one exciting the small sphere and the other the large sphere.

Figure 5(a) presents a schematic and bright field image of a three-parallel microspheres resonator configuration. In Figure 5(b), we depict the electric field intensity distribution for a three-parallel microsphere resonator with the same diameter at a wavelength of 675 nm. The simulation results reveal a highly intensified electric field in the first microsphere resonator, which decays as it is coupled into the second and third microspheres. When light enters the first microsphere resonator, it undergoes multiple reflections along the inner surface, propagating along the circumference of the sphere through total internal reflection. However, as the light propagates from the first microsphere into the second and third microspheres, it encounters interfaces between the spheres. At these interfaces, a portion of the incident light intensity is reflected back. Additionally, as light propagates through the system, it also experiences scattering and absorption losses, further contributing to the decrease of light intensity in the second and third microspheres.

Figures 5(c)-5(e) display the relative intensity of the PL spectra within the wavelength range of 640–760 nm for the amplifier incorporating three parallel microsphere resonators. The experiments were conducted at four different laser power levels: 0.1 mW, 0.2 mW, 1 mW, and 2 mW. Similar to the previous experiments with two microsphere resonators, we observe that at the lower laser power levels of 0.1 mW and 0.2 mW, the number of WGM peaks is larger compared to the higher laser power levels of 1 mW and 2 mW. At the two lower laser power levels, 18 short peaks emerge, with Q-factors ranging from 100 to 500, excited across a wide range of wavelengths corresponding to various WGMs. For the higher laser power levels of 2 mW and 1 mW, distinct strong WGM peaks appear at 675 nm, with Q-factors of 1380 and 1100, respectively. The adjacent peaks observed in this region are relatively short, except for a WGM peak at 677 nm, which has Q-factors of 670 and 130 for laser power levels of 2 mW and 1 mW, respectively.

Figure 6 shows the results obtained for a triangular configuration of three microsphere resonators. In Fig. 6(a), we show a schematic and bright field image of the three microsphere resonators in this configuration. Figure 6(b) shows the electric field intensity distribution for the same sizes and configuration at the wavelength of 675 nm. The simulated electric field intensity distribution reveals a highly intensified electric field in the lower left microsphere resonator, while the light intensity decays equally as it couples into the two adjacent microspheres. Figures 6(c)-(e) show the PL spectra for the triangular configuration of three-microsphere resonators when the laser power level is gradually increased to 2 mW.

For the lower values of excitation power, the WGM structure is clearly present, but the number of WGM peaks decreases with the increase in laser power. As the laser excitation power is increased, the undulations resulting from the WGMs decrease in intensity, while the spectra become predominantly characterized by a narrow and dominant WGM peak at the wavelength of 675 nm. The Q-factors of these WGMs are measured to be 675, 2400, and 3750 for laser power levels of 0.2 mW, 1 mW, and 2 mW, respectively. These results demonstrate that the Q-factors of the triangular configuration of three microsphere resonators are ~2.2 and ~2.7 times larger than those of the parallel arrangement for the same laser power levels of 1 mW and 2 mW, respectively. Furthermore, the triangular arrangement exhibits a greater number of WGMs compared to the parallel microsphere resonator. At the lower laser power level of 0.2 mW, 26 short peaks of relative intensity with Q-factors ranging from 100 to 500 emerge in the triangular configuration of three microsphere resonators, compared to 18 short WGM peaks observed in the parallel configuration. Moreover, the average peak-to-peak distance along the wavelength axis of the WGM peaks at a laser power of 0.2 mW is ~6 nm for the parallel configuration, while it is reduced to ~4 nm for the triangular configuration. These findings highlight the influence of the microsphere arrangement on the optical properties of the resonators, with the triangular configuration exhibiting enhanced Q-factors, a higher number of WGMs, and a more compact distribution of WGM peaks compared to the parallel microsphere resonator.

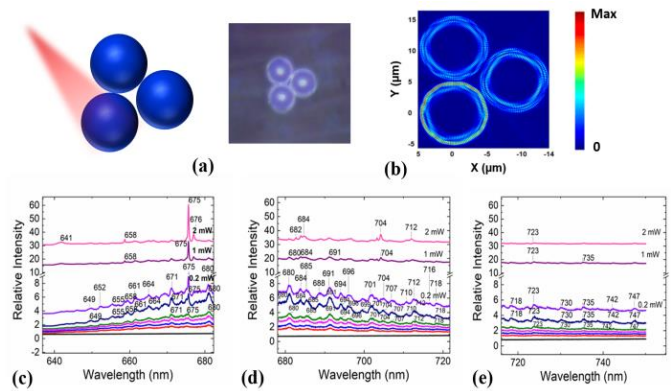

Figure 6. (a) Schematic and bright field image for three microsphere resonators in triangular configuaration. (b) Electric field intensity calculated for three microsphere resonators in triangular configuration at λ = 675 nm plotted in the xy plane (top view of the spheres). (c-e) PL spectra in the 640–750 nm wavelength range for an amplifier incorporating three microsphere resonators in triangular configuration for various laser power levels.

Figure 7 showcases three microsphere resonators of equal sizes with laser excitation of different polarizations. Figure 7(a) shows the bright field image of the three microsphere resonators for parallel polarization. Figure 7(b) shows the PL spectra of the three microsphere resonators for parallel polarization and excitation at the right, middle, and left spheres within the 640–750 nm wavelength range. It is observed that the location of excitation significantly affects the number and intensity of WGM peaks for the same laser power level of 0.2 mW. Excitation of the left and right spheres results in 29 and 27 short WGM peaks, respectively. In contrast, excitation of the middle sphere leads to relatively flat PL spectra, suppressing most other WGM modes. The dominant WGM peaks are observed at the resonance wavelength of 676 nm for all three cases of parallel polarization. The Q-factors at this wavelength are measured to be 1680, 600, and 800 for laser excitation at the middle, left, and right spheres, respectively.

Figure 7(c) shows the bright field image of the three-microsphere resonator for different polarizations when the right sphere is excited, while the laser power level is kept constant in all measurements at 0.2 mW. In Figure 7(d), the PL spectra of this resonator are shown for circular, orthogonal, and parallel polarization when the right sphere is excited. The number of WGM peaks observed in the PL spectra varies depending on the polarization. Parallel polarization exhibits more WGM peaks compared to the other two polarizations. The dominant WGM peaks are observed at the resonance wavelength of 676 nm for all three cases of circular, orthogonal, and parallel polarization. The Q-factors at this wavelength are measured to be 1690, 750, and 890 for circular, orthogonal, and parallel polarization, respectively. The higher Q-factor for circular polarization implies that the resonant modes excited under this configuration exhibit higher energy confinement and lower energy losses. Conversely, orthogonal and parallel polarizations demonstrate smaller Q-factors, indicating higher energy losses compared to circular polarization. The polarization of the incident light affects the excitation efficiency of different modes and influences the energy distribution within the microspheres. As a result, different polarization configurations can lead to variations in the observed PL spectra, WGM peaks, and Q-factors.

Figure 7(e) shows the bright field image of the three microsphere resonators when the laser is focused on the junction of the right and middle spheres. In Fig. 7(f), the PL spectra are shown when the laser with a power level of 0.2 mW is focused on the junction of the right and middle spheres. The number and intensity of WGM peaks in the PL spectra are found to be smaller for orthogonal polarization compared to circular and parallel polarizations. The dominant WGM peaks are observed at the resonance wavelength of 675 nm for all three cases of circular, orthogonal, and parallel polarization. The Q-factors at this wavelength are measured to be 2600, 2200, and 2850 for circular, orthogonal, and parallel polarization, respectively. Interestingly, the Q-factors for circular, orthogonal, and parallel polarizations differ when the laser is focused on the junction of the right and middle spheres compared to when it is focused solely on the right sphere. For the laser focused on the junction, the Q-factors are 1.5, 2.9, and 3.2 times higher compared to circular, orthogonal, and parallel polarizations, respectively. The observed differences in the number and intensity of WGM peaks and the Q-factors can be attributed to the specific geometry and mode coupling effects between the spheres. Focusing the laser on the junction of the right and middle spheres likely introduces additional complexity in the mode excitation and coupling behavior. This can result in enhanced energy confinement and more efficient excitation of resonant modes, leading to higher Q-factors and a larger number of observed WGM peaks for circular and parallel polarizations compared to orthogonal polarization. These findings highlight the influence of polarization and geometry on the excitation and behavior of WGMs in microsphere resonators.

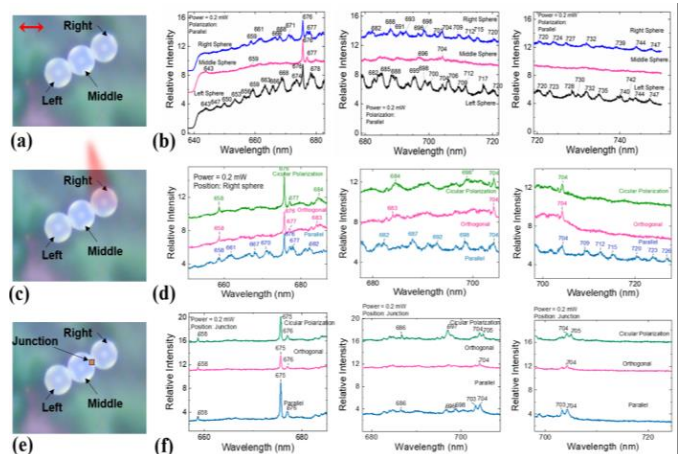

Figure 7. (a) Bright-field image for three microsphere resonators with equal sizes for parallel polarization. (b) PL spectra of three microsphere resonators for parallel polarization and excitation at the right, middle, and left spheres within the 640–750 nm wavelength range. (c) Bright-field image of three microsphere resonators for parallel polarization when the right sphere is excited. (d) PL spectra of three microsphere resonators for circular, orthogonal, and parallel polarization when the right sphere is excited. (e) Bright-field image of three microsphere resonators when the laser is focused on the junction of the right and middle spheres. (f) PL spectra of three microsphere resonators for circular, orthogonal, and parallel polarization, when the laser is focused on the junction of the right and middle spheres. The laser power level was kept the same for all the measurements at 0.2 mW.

In addition, a three-microsphere resonator system of equal sphere sizes was investigated with laser excitation at various locations (top edge, right edge, middle, bottom edge, and left junction point of the right sphere) for three different polarizations (parallel, orthogonal, and circular). Figure 8(a) shows the bright field image of three microsphere resonators for parallel polarization when these five locations of the right microsphere are excited by the laser with a power level of 0.2 mW. For parallel polarization, the location of excitation significantly affects the number and intensity of WGM peaks in the PL spectra as shown in Figure 8(b). Excitation at the middle point resulted in a larger number of WGM peaks across the 640-750 nm wavelength range, indicating efficient excitation of a broader range of resonant modes. On the other hand, excitation at the bottom point suppressed WGM peaks, suggesting inefficient coupling to the resonant modes. The dominant WGM peaks observed at the resonance wavelength of 676 nm for the other four excitation locations indicate that this is a favored resonant mode with high Q-factors for the microspheres under various excitation conditions. For parallel polarization, the Q-factors are 2850, 2100, 900, and 700 for laser excitation at the left junction edge, right edge, top edge, and middle point of the right sphere, respectively. In other words, the Q-factors vary depending on the excitation location, suggesting differences in the energy confinement and loss mechanisms for each excitation configuration. The excitation location affects how the incident light interacts with the resonant modes and influences the resulting PL spectra and observed WGM peaks.

Figure 8(c) represents the bright field image of three microsphere resonators for orthogonal polarization when the five locations on the edges and middle point of microsphere are excited by the same laser power level of 0.2 mW. Figure 8(d) shows the PL spectra of the three microsphere resonators for orthogonal polarization and various excitation locations on the right sphere. Similar trends were observed for orthogonal polarization. Excitation at the bottom point resulted in suppressed WGM peaks, but the number of peaks was smaller compared to parallel polarization, indicating fewer efficiently excited resonant modes. However, the number of WGM peaks for excitation at the bottom point are not large for orthogonal polarization compared to the number of WGM peaks for parallel polarization, indicating that a smaller number of resonant modes are efficiently excited under this polarization and excitation configuration. For orthogonal polarization, the Q-factors at a resonance wavelength of 676 nm are 2200, 1850, 900, and 750 for laser excitation at the left junction edge, right edge, middle point, and top edge of the right sphere, respectively.

Figure 8(e) represents the bright field image of the same three microsphere resonators for circular polarization when the four edges and middle point of the microsphere are excited. For circular polarization, excitation at the bottom point also suppressed WGM peaks, and the observed behavior was similar to that of parallel and orthogonal polarizations, as shown in Fig. 8(d). Thus, when the laser is focused on the bottom point of the microsphere, the WGM modes are independent of polarization types, and thereby the excitation at this location does not efficiently couple to the resonant modes of the microsphere. For circular polarization, the Q-factors at a resonance wavelength of 676 nm were 2600, 2000, 1900, and 800 for laser excitation at the left junction edge, right edge, middle point, and top edge of the right sphere, respectively. The excitation location affects how the incident light interacts with the resonant modes and alters the energy confinement and loss mechanisms, which results in different intensities of WGM peaks observed in the PL spectra.

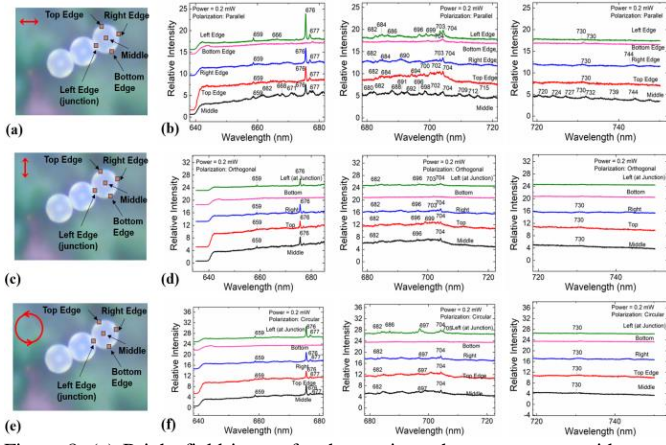

Figure 8. (a) Bright-field image for three microsphere resonators with equal sizes for parallel polarization with laser excitation at five locations on the right sphere: top edge, right edge, middle, bottom edge, and left junction point. (b) PL spectra of three microsphere resonators for parallel polarization and different excitation locations within the 640–750 nm wavelength range. (c) Bright-field image of three microsphere resonators for orthogonal polarization when the right sphere is excited at five different locations. (d) PL spectra of three microsphere resonators for orthogonal polarization for different excitation locations on the right sphere. (e) Bright-field image of three microsphere resonators for circular polarization when the right sphere is excited at five different locations. (f) PL spectra of three microsphere resonators for circular polarization for different excitation locations on the right sphere.

These findings highlight the influence of polarization (circular, orthogonal, or parallel) with different laser excitation locations on the right microsphere. In the case where the laser is excited at the junction of the microsphere, it is observed that the parallel polarization results in a larger WGM peak compared to circular and orthogonal polarizations. This indicates that the resonator efficiently couples to and supports a stronger WGM at the specified resonance wavelength under parallel polarization. However, when the laser is excited at the middle point of the microsphere, the circular polarization exhibits a stronger WGM peak at the resonance wavelength of 676 nm compared to orthogonal and parallel polarizations. This finding reveals that circular polarization is more effective in exciting and confining the resonant mode at the middle location, resulting in a stronger WGM peak in the PL spectra. The differences in the observed behavior can be attributed to polarization-dependent excitation and mode coupling effects within the microsphere resonator, which affect how light interacts with the resonant modes, resulting in WGM peaks with different intensities at a specific resonance wavelength.

In addition, results for a system of five microsphere resonators with equal sizes were compared with a configuration of three microsphere resonators arranged in a triangular configuration (Figure 9). Figure 9(a) shows the schematic and bright field image of the five microsphere resonators with the middle sphere excited by the laser. Figure 9(b) shows a schematic of the five microsphere resonators with the top sphere excited by the laser. Figure 9(c) shows this configuration's electric field intensity distribution at the wavelength of 676 nm. The simulation results indicate a highly intensified electric field in the microsphere resonator excited by the laser, but the intensity decays as light is coupled into subsequent microspheres, particularly in the last two microsphere resonators.

Figures 9(d)-9(f) show the PL spectra for the five microsphere resonators with two different laser excitation locations on the first and middle microspheres. They also show PL spectra for three microsphere resonators in a triangular configuration. It is observed that the PL spectra of the three microsphere resonators in triangular configuration is similar to that of the five microsphere resonators with laser excitation on the top microsphere. The number and intensity of WGM peaks in the PL spectra of the five microsphere resonators with laser excitation on the middle microsphere are comparable to those of the three microsphere resonators in the triangular configuration, but smaller than those of the five microsphere resonators with laser excitation on the top microsphere. The comparison of corresponding TE and TM mode numbers are shown in Figure S3 and Table S1. The dominant WGM peak for the five microsphere resonators with laser excitation on the middle microsphere still occurs at the resonance wavelength of 676 nm, with a Q-factor of 1700. Conversely, the dominant WGM peak for the laser excitation on the top microsphere is observed at a resonance wavelength of 691 nm, with a Q-factor of 1800. These findings demonstrate that the arrangement and location of laser excitation significantly influence the number, intensity, and characteristics of WGM peaks in the microsphere resonator system.

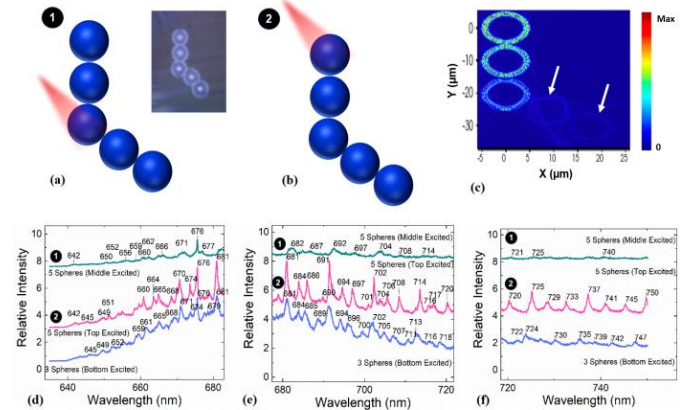

Figure 9. (a) Schematic and bright field images of five microsphere resonators with equal sphere sizes and laser excitation at the middle sphere. (b) Schematic of five microspheres with the top sphere excited. (c) Electric field intensity calculated for the five microsphere resonators excited at the top sphere at $\lambda$ = 676 nm and plotted in the xy plane (top view of the spheres). (d-f) PL spectra in the 650–750 nm wavelength range for the five microspheres where top and middle spheres are excited. PL spectra are also shown for three microsphere resonators in a triangular configuration.

We also investigate a nine-microsphere resonator combination, where the centered (fifth) microsphere has a larger diameter of 15 µm compared to the other microspheres with diameters of 10 µm. Figure 10 shows the results for this configuration. Figure 10(a) shows a schematic and bright field image of the nine-microsphere resonator, showcasing the arrangement of the microspheres. The electric field intensity distribution at the wavelength of 679 nm is displayed in Fig. 10(b). The results reveal a highly intensified electric field in the first microsphere resonator excited by the laser, while the intensity gradually decays as light is coupled into subsequent microspheres. Despite the long chain of microspheres in this experiment, coupling between them is still visible in the field distribution, indicating the excitation of WGMs in all microspheres. Figure 10(c) shows the PL spectra for the nine-microsphere resonator configuration for four different laser power levels: 0.02, 0.1, 0.2, and 1 mW. It is observed that the number of WGM peaks is small across all laser power levels. However, the intensity of the dominant WGM peak increases with higher laser power at the resonance wavelength of 675 nm. The Q-factors of these WGM peaks are measured to be 1100 and 400 for laser power levels of 1 mW and 0.2 mW, respectively.

These findings demonstrate the behavior of the nine-microsphere resonator configuration and highlight the excitation of WGMs, even in longer chains of microspheres. The PL spectra indicate a dependence on the laser power level, with higher power levels leading to an increase in the intensity of the dominant WGM peaks. These results provide valuable insights into the optical properties and behavior of WGMs in complex microsphere resonator configurations.

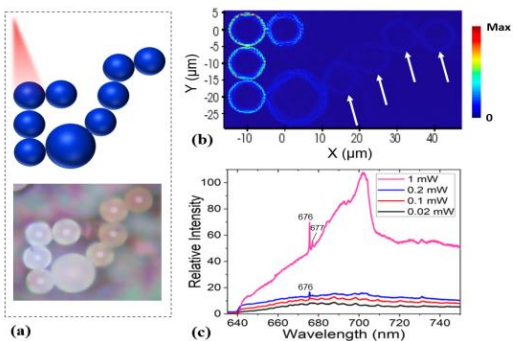

Figure 10. (a) Schematic and bright field image of nine microsphere resonators. (b) Electric field intensity calculated for nine microsphere resonators with unequal sphere sizes at λ = 679 nm and plotted in the xy plane (top view of the spheres). (c) PL spectra of the nine microsphere resonators within the 650–750 nm wavelength range for four different laser power levels of 0.02, 0.1, 0.2, and 1 mW.

## IV. POTENTIAL APPLICATIONS IN SENSING AND INFORMATION PROCESSING

Our findings introduce an optoplasmonic amplifier, harnessing injection seeding through an internally-generated Raman signal, and operating in the visible wavelength range. The presented configurations comprise a gain medium of fluorescence molecules, coated to the outside of spheres, intricately associated with a WGM microsphere resonator 52. These optical systems are designed to meticulously amplify a single, or a limited number of Raman lines generated within the WGM-based microsphere resonators, making it particularly well-suited for routing narrowband optical power on an integrated chip. Our approach leverages injection-locking to achieve precise spectral control and tailor the effective Q-factor of the amplifier, a departure from previous WGM-based devices where Q was solely determined by the resonator. In our setup, the resonator is positioned within the evanescent optical field of the microsphere, allowing specific WGMs, associated with the spherical resonator, efficiently harvest energy from fluorescence molecules with power circulating in the microsphere resonator 52. This work illustrates the capability to generate an internal Raman signal via injection seeding, aligning the Raman seed radiation precisely with a specific resonator mode that provides exceptional control over the output spectrum. By carefully configuring this amplifier, we can predominantly extract energy stored under the amplifier gain profile at a single resonance wavelength 56.

Achieving precise light manipulation at micro- and nano-spatial scales is a prerequisite for advancing biosensing applications, including the integration of biomedical sensors on-chip. The versatility of Q values achievable with different configurations of microsphere resonators enables the observation of nonlinear optical processes at field intensities orders of magnitude lower than those required by bulk nonlinear media 57. The diminutive mode volumes characteristic of WGM-based microsphere resonators at high laser power levels result in extended photon cycle times within the cavity, leading to narrower resonant wavelength lines, intensified energy density, and augmented interactions between light and the surrounding environment. These attributes hold significant promise for a range of sensing applications. Notably, the high energy density and narrow resonant wavelengths associated with WGM microsphere resonators enable precision in detecting even the slightest variations in the surrounding environment. This results in enhanced sensitivities and lower detection limits compared to conventional sensors.

Furthermore, the compound photonic-plasmonic device described herein presents an optical system ideally suited for parallel, distributed systems geared towards the storage, amplification, and routing of optical power within an optical system on a chip 52, 58. The high intensity of WGMs allows for stronger and more robust optical signals, which could enable efficient transmission and processing of data on a chip. The intensity of WGMs determines the strength of the optical signal within the resonator, and stronger signals can provide more reliable and accurate information processing. By amplifying specific WGMs, the network can prioritize and direct signals to specific regions or nodes, enhancing connectivity and facilitating targeted information processing. Nonlinearities and amplification of WGM microresonators enable efficient interconnectivity and information flow encoded on optical signals, improving the overall efficiency and the noise resiliency of optical signals in future photonic integrated circuits on a chip. Stronger and amplified WGMs can enhance the resilience of these circuits to noise and signal degradation. In practical scenarios, optical signals can be subjected to various forms of noise and attenuation. By utilizing highly intense and amplified WGMs, the photonic integrated circuits can mitigate the adverse effects of noise, ensuring accurate and reliable data processing. WGMs could therefore play a vital role in future information processing with light by providing stronger signals, enabling nonlinear computations, facilitating interconnectivity, and improving noise resilience. These factors collectively contribute to the overall performance, efficiency, and ability to perform complex optical information processing tasks.

## V. Summary and Conclusion

This study examined the optical characteristics of microsphere resonators in diverse setups: single spheres, double spheres with equal or unequal sphere sizes, parallel and triangular arrangements of three spheres, as well as configurations with five and nine microspheres. The impact of excitation location, polarization, microsphere size, and arrangement on the resonance peaks and Q-factors was thoroughly investigated. For single microsphere resonators, it was observed that WGMs exhibited a highly intensified electric field at λ = 679 nm, resulting in improved sensitivity and detection capabilities. The PL spectra exhibited distinct dominant peaks at higher laser power levels, indicating enhanced optical signals which could lead to efficient sensing and data processing in optical systems based on such structures. In the case of double microsphere resonators, the PL spectra exhibited multiple short peaks at lower laser power levels, while only one dominant peak was observed at higher power levels. Double microsphere resonators with unequal sphere sizes demonstrated variations in the intensity and the number of WGM peaks, depending on the excitation location and sphere sizes. FDTD simulations for three or more parallel microsphere resonators demonstrated strong standing electromagnetic waves in the excited microsphere and decay in the electric field intensity as light coupled from the first into the subsequent microspheres.

Similarly, the PL spectra exhibited larger WGM peaks at lower laser power levels, with distinct peaks at higher power levels. In the triangular configuration of three microsphere resonators, the PL spectra had fewer WGM peaks compared to the parallel configuration. The Q-factors for the triangular arrangement were higher, indicating improved energy confinement and a more compact distribution of WGM peaks. The excitation location and polarization of the incident light significantly affected the number and intensity of WGM peaks in the three-microsphere resonator configuration. Circular polarization led to higher Q-factors, while excitation at specific locations of a microsphere resulted in varied peak intensities and distributions. The results show that the arrangement and location of laser excitation significantly influence the number, intensity, and characteristics of WGM peaks in the microsphere resonator system. Finally, the nine-microsphere resonator configuration demonstrated that even in longer chains of microspheres, the PL spectra exhibited increased intensity of the dominant WGM peaks with higher laser power levels suppressing other WGM peaks across a wide range of wavelengths. Overall, this work contributes to the understanding WGM behavior in microsphere resonator structures, highlighting their potential for sensing and optical neural network applications. The influence of configuration, size, excitation, and polarization on WGM properties provides valuable insights for designing and optimizing microsphere resonator systems.

## VI. Acknowledgments

Manas Ranjan Gartia was supported by National Science Foundation (NSF CAREER award number: 2045640).

52. Gartia, M. R.; Seo, S.; Kim, J.; Chang, T.-W.; Bahl, G.; Lu, M.; Liu, G. L.; Eden, J. G., Injection-seeded optoplasmonic amplifier in the visible. *Scientific reports* **2014**, *4* (1), 6168.
53. Wenzel, H.; Schwertfeger, S.; Klehr, A.; Jedrzejczyk, D.; Hoffmann, T.; Erbert, G., High peak power optical pulses generated with a monolithic master-oscillator power amplifier. *Optics Letters* **2012**, *37* (11), 1826-1828.
54. Li, Y.; Svitelskiy, O.; Maslov, A.; Carnegie, D.; Rafailov, E.; Astratov, V. In *Giant resonant light forces in microspherical photonics, Light: Sci*, Appl: 2013.
55. Li, Y. Microspherical photonics: Giant resonant light forces, spectrally resolved optical manipulation, and coupled modes of microcavity arrays. The University of North Carolina at Charlotte, 2015.
56. Eden, J. G.; Gartia, M. R.; Liu, G. L., Injection-seeded whispering gallery mode optical amplifier devices and networks. Google Patents: 2018.
57. Lin, H.-B.; Campillo, A., CW nonlinear optics in droplet microcavities displaying enhanced gain. *Physical review letters* **1994**, *73* (18), 2440.
58. Liu, J.; Wu, Q.; Sui, X.; Chen, Q.; Gu, G.; Wang, L.; Li, S., Research progress in optical neural networks: theory, applications and developments. *PhotoniX* **2021**, *2* (1), 1-39.

## SUPPORTING INFORMATION

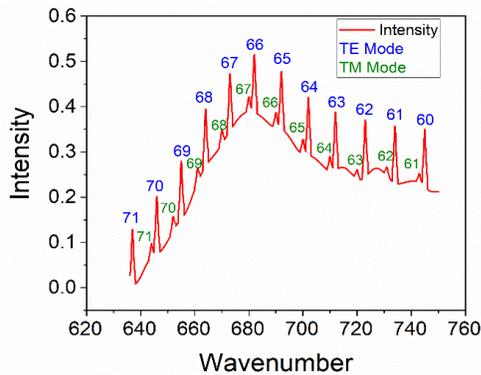

**Figure S1**: Intensity vs. wavelength, indicating the values of TE and TM modes.

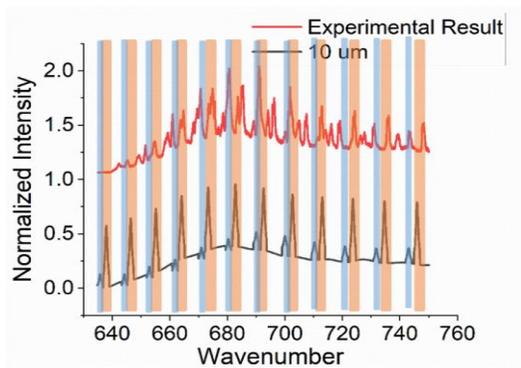

**Figure S2**: Normalized Intensity vs. wavelength comparing the experimental result and the simulation validation for 10 μm diameter of microsphere. TE = brown color; TM = blue color.

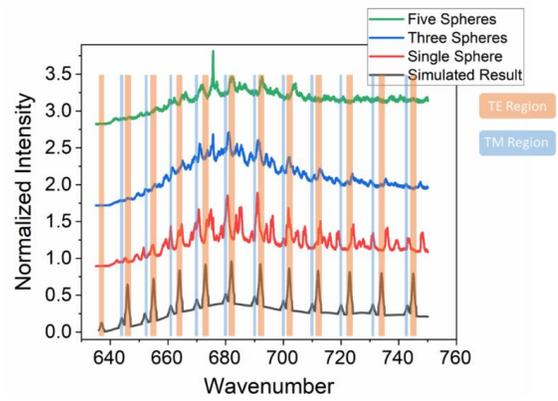

**Figure S3**: Normalized Intensity vs. wavelength for single, three, and five spheres, indicating the intensity peaks in TE and TM regions. TE = brown color; TM = blue color.

**Table S1.** Wavelength positions of TM and TE modes correspond to different microsphere diameters.

| TM Mode | | | | | TE Mode | | | | |
|---|---|---|---|---|---|---|---|---|---|
| Diameter of the Sphere | | | | | | | | | |
| 10.00 μm | 10.01 μm | 9.99 μm | 10.05 μm | 9.95 μm | 10.00 μm | 10.01 μm | 9.99 μm | 10.05 μm | 9.95 μm |
| Wavelength positions of the mode | | | | | | | | | |
| 754.60 | 755.36 | 753.85 | 758.38 | 750.83 | 745.96 | 746.71 | 745.22 | 749.69 | 742.23 |
| 743.01 | 743.75 | 742.26 | 746.72 | 739.29 | 734.63 | 735.36 | 733.89 | 738.30 | 730.95 |
| 731.77 | 732.50 | 731.03 | 735.43 | 728.11 | 723.64 | 724.36 | 722.91 | 727.25 | 720.02 |
| 720.87 | 721.59 | 720.14 | 724.47 | 717.26 | 712.97 | 713.68 | 712.26 | 716.54 | 709.41 |
| 710.29 | 711.00 | 709.58 | 713.84 | 706.74 | 702.62 | 703.32 | 701.92 | 706.14 | 699.11 |
| 700.02 | 700.72 | 699.32 | 703.52 | 696.52 | 692.57 | 693.26 | 691.88 | 696.04 | 689.11 |
| 690.05 | 690.74 | 689.36 | 693.50 | 686.60 | 682.81 | 683.49 | 682.13 | 686.22 | 679.40 |
| 680.36 | 681.04 | 679.68 | 683.76 | 676.96 | 673.32 | 674.00 | 672.65 | 676.69 | 669.96 |
| 670.94 | 671.62 | 670.27 | 674.30 | 667.59 | 664.1 | 664.76 | 663.43 | 667.42 | 660.78 |
| 661.79 | 662.45 | 661.13 | 665.10 | 658.48 | 655.13 | 655.78 | 654.47 | 658.40 | 651.85 |
| 652.88 | 653.53 | 652.23 | 656.15 | 649.62 | 646.4 | 647.04 | 645.75 | 649.63 | 643.16 |
| 644.21 | 644.86 | 643.57 | 647.44 | 640.99 | 637.89 | 638.54 | 637.26 | 641.09 | 634.71 |
| 635.78 | 636.41 | 635.14 | 638.96 | 632.60 | 629.62 | 630.25 | 628.99 | 632.77 | 626.48 |